%% file: main.tex
\documentclass{article}

\usepackage[dblblindworkshop, final]{neurips_2026}

\usepackage[utf8]{inputenc}
\usepackage[T1]{fontenc}
\usepackage{hyperref}
\usepackage{url}
\usepackage{booktabs}
\usepackage{amsfonts}
\usepackage{graphicx}
\usepackage{microtype}
\usepackage{tikz}
\usetikzlibrary{arrows.meta,positioning}

\title{AgentScript-Eval: Benchmarking LLMs and Agents for Code Generation in an Enterprise DSL}

\workshoptitle{2nd Workshop on Agentic AI Benchmarks and Applications
for Enterprise Tasks (AABA4ET), NeurIPS 2026}

\author{%
  Dipin Khati$^{1,2}$\thanks{Work done during an internship at Salesforce.}\quad
  Shubham Mehrotra$^{2}$\quad
  Bin Bi$^{2}$\quad
  Zhou Yu$^{2}$\quad
  Denys Poshyvanyk$^{1}$\\
  \textbf{James Zhu}$^{2}$\quad
  \textbf{Sitaram Asur}$^{2}$\quad
  \textbf{Phil L. Mui}$^{2}$\\
  $^{1}$William \& Mary\quad
  $^{2}$Salesforce\\
  \texttt{dkhati@wm.edu}
}

\begin{document}
\maketitle

\input{sections/00_abstract}
\input{sections/01_introduction}
\input{sections/02_agentscript_and_benchmark}
\input{sections/03_evaluation}

\input{sections/04_discussion}
\input{sections/05_related_work}

\bibliographystyle{plainnat}
\bibliography{references}

\appendix
\input{sections/06_appendix}

\end{document}

%% file: sections/00_abstract.tex
\begin{abstract}
Domain-specific languages (DSLs) are widely used in enterprise software, giving a compact interface to complex tasks, from data access to infrastructure automation. Platforms increasingly extend them to agents, where a developer declares the topics, permitted actions, and guardrails of an agent rather than writing general-purpose code. Large language models (LLMs) write general-purpose code well, but that ability has been measured almost entirely on languages abundant in public pretraining data, and enterprise DSLs are neither abundant nor public. We introduce
\textbf{AgentScript-Eval}, an execution-graded benchmark of $100$ real tasks for
AgentScript, the declarative DSL that Salesforce uses in Agentforce to
build conversational agents. Tasks span authoring an agent from a specification,
debugging a broken definition, and optimizing a working one.
Evaluating five models zero-shot, we find that the regime is hard: the best model reaches
only $25.0\%$ pass@3, and every model scores $0\%$ on authoring from scratch. We
further find that model scale and price are weak predictors of success---a $31$B
open-weight model attains the highest compilation rate of any model tested ($48.0\%$)
and matches the strongest frontier model on partial-credit accuracy at the lowest
token cost. Adding an agentic compile-and-edit loop is the single largest lever,
roughly doubling partial-credit accuracy. Beyond that gain, the wording of domain
guidance matters more than its packaging. The same skill pack scores $28\%$ as plain
documentation but only $14\%$ as a live plugin. Rewriting one skill inside that plugin
then raises it to $37\%$, our best result, with catalog and tooling untouched.
A per-category breakdown shows the difficulty
is concentrated in authoring from scratch, where no zero-shot model succeeds, and even
the best agentic configuration stays under $17\%$. Larger models did not reliably outperform smaller ones; what mattered was the harness surrounding the model and the guidance supplied to it.
\end{abstract}

%% file: sections/01_introduction.tex
\section{Introduction}
\label{sec:intro}

Code generation is among the most extensively benchmarked capabilities of LLMs, and 
by prevailing measures it appears close to saturation: frontier systems exceed $90\%$ on HumanEval-style 
function synthesis~\citep{chen2021evaluating,evalplus-leaderboard} and, embedded in agentic harnesses, 
resolve a substantial share of real GitHub issues~\citep{jimenez2023swe,yang2024sweagent,swebench-leaderboard}. 
These results rest on an assumption that is rarely made explicit: that the target language is well represented in 
pretraining data. Python and Java dominate existing benchmarks in part because they are also among the languages 
most abundantly represented in the public code corpora on which these models are trained. Thus, benchmark 
performance may overstate the generality of LLM code-generation capabilities across programming languages.

This limitation is particularly consequential in enterprise software, where substantial portions of production systems are built using domain-specific and declarative languages rather than general-purpose programming languages. Conversational agents, through which customers and employees interact with organizations, are authored on 
major enterprise platforms through declarative surfaces rather than general-purpose 
code~\citep{copilotstudio-topics,dialogflow-cx-flows,lex-v2-intents}. These surfaces let developers specify 
an agent's behavior, including what it may discuss, which actions it may invoke, and which guardrails 
constrain it, while the platform handles execution. Such DSLs deliberately constrain the programming 
interface to expose abstractions tailored to a particular domain~\citep{mernik2005dsl}.

Precisely because these languages are specialized and closely tied to particular enterprise platforms, they are also among the settings in which least is known about LLM code generation. They are scarce or absent from public pretraining corpora, evolve alongside their underlying platforms rather than through broadly adopted standards, and lack established public benchmarks. Existing DSL research offers limited coverage of this setting, targeting languages that are themselves public and well represented, such as SQL~\citep{yu2018spider} and IT-automation-YAML~\citep{pujar2023ansible}, while studies of resource scarcity vary the language only across general-purpose targets~\citep{cassano2022multipl}. We are aware of no publicly executed benchmark for an enterprise-specific DSL used in active production. Whether the competence documented above transfers to such languages, and what can restore it when it does not, therefore remains an open question.

We address this question in the context of one such language. \textbf{AgentScript} is a 
declarative DSL used in Salesforce Agentforce to author conversational 
agents~\citep{agentscript-docs}. We introduce \textbf{AgentScript-Eval}, a benchmark of $100$ tasks 
drawn from real agent-development work and spanning the developer lifecycle: authoring an agent from 
a specification, repairing a broken agent, and optimizing a working one. Each task is graded by compiling 
the generated program and executing it against multi-turn acceptance checks, distinguishing programs 
that parse but misbehave from those that fail to compile. We evaluate five models across seven generation 
configurations, ranging from single-shot generation to agentic loops with and without domain-skill-packs.

Our contributions are threefold. First, we introduce AgentScript-Eval, to our knowledge the first verifiable 
benchmark for enterprise-specific, in-production agent-authoring DSL (Section~\ref{sec:benchmark}). Second, we provide a 
controlled comparison across models and generation configurations that separates compilation from behavioral 
correctness and reports token cost per task (Section~\ref{sec:eval}). Third, we identify three findings that complicate 
the extrapolation of general-purpose coding. The regime is substantially harder than the existing results would suggest: 
the strongest model achieves only $25.0\%$ pass@3, and no model solves an authoring task zero-shot. 
Scale and price are weak predictors: a $31$B open-weight model compiles more programs than any hosted model 
we evaluate, at the lowest cost per task. Finally, once a model can compile and edit its own output, 
the \emph{wording} of domain guidance matters more than its \emph{packaging}: rewriting a single skill 
within an otherwise unchanged plugin yields a $+23$ point improvement and our strongest configuration at $37\%$.

%% file: sections/02_agentscript_and_benchmark.tex
\section{Method}
\label{sec:benchmark}
\label{sec:method}

\paragraph{AgentScript.} AgentScript is a declarative DSL for authoring
conversational agents in Salesforce Agentforce~\citep{agentscript-docs}:
a program specifies the agent's topics, the actions it may invoke with their inputs and
outputs, and guardrails on behavior (Appendix~\ref{app:language}). Actively developed and
absent from public corpora, it demands both valid syntax and behavior matching a
specification.

\paragraph{Tasks.} AgentScript-Eval contains $100$ real agent-development tasks: \textbf{authoring} ($30$) produces an agent from a natural-language specification, \textbf{debugging} ($40$) repairs an uncompiling AgentScript, and \textbf{optimization} ($30$) improves a working one against a stated objective. Authoring is most demanding, since debugging and optimization supply an existing program as scaffold. Each task ships with an input context and executable checks; the defect taxonomy and construction details are in Appendix~\ref{app:construction}.

\begin{figure}[!tb]
\centering
\begin{tikzpicture}[
  font=\footnotesize,
  base/.style={draw=black!55, semithick, rounded corners=2.5pt, align=center,
               inner xsep=4.5pt, inner ysep=2pt, minimum height=6.5mm},
  io/.style  ={base, fill=black!6},
  proc/.style={base, fill=blue!7},
  eval/.style={base, fill=teal!9},
  arr/.style ={-{Latex[length=1.7mm]}, semithick},
  fb/.style  ={-{Latex[length=1.7mm]}, dashed, semithick, blue!55!black},
  node distance=3.8mm
]
\node[io]                  (task) {\textbf{Task}\\[-1.5pt]{\scriptsize\color{black!60}spec\,+\,program}};
\node[proc, right=of task] (gen)  {\textbf{Generate}\\[-1.5pt]{\scriptsize\color{black!60}direct or agentic}};
\node[proc, right=of gen]  (comp) {\textbf{Compile}\\[-1.5pt]{\scriptsize\color{black!60}compile rate}};
\node[eval, right=of comp] (exec) {\textbf{Execute}\\[-1.5pt]{\scriptsize\color{black!60}stubbed dialogues}};
\node[eval, right=of exec] (met)  {\textbf{Score}\\[-1.5pt]{\scriptsize\color{black!60}pass rate, pass@$k$}};

\draw[arr] (task) -- (gen);
\draw[arr] (gen)  -- (comp);
\draw[arr] (comp) -- (exec);
\draw[arr] (exec) -- (met);

\draw[fb] (comp.south) -- ++(0,-2.2mm) -|
  node[pos=0.25, below, font=\scriptsize, color=blue!55!black]
  {compile feedback (agentic)} (gen.south);

\draw[fb] (met.north) -- ++(0,2.2mm) -|
  node[pos=0.22, above, font=\scriptsize, color=blue!55!black]
  {execution feedback (agentic)} (gen.north);
\end{tikzpicture}
\caption{The evaluation harness. All configurations share the same tasks, environment,
and checks, differing only in how the program is produced}
\label{fig:pipeline}
\end{figure}
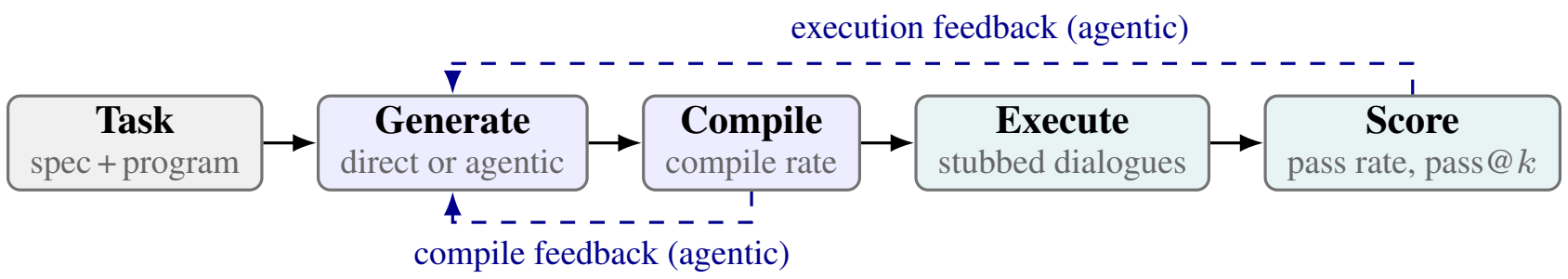

\paragraph{Grading.} Text-similarity metrics correlate poorly with functional
correctness for code~\citep{evtikhiev2022bleu,liu2023your}, so we grade by compilation
and execution. If the generated agent
compiles, it is executed as an agent against the task's acceptance checks, which
encode expected actions, state, and responses over multi-turn dialogues with stubbed
backends.

\paragraph{Metrics.} We report \textbf{compile rate}, the fraction of tasks whose
program compiles; \textbf{pass rate}, the mean per-task fraction of checks that succeed
(partial credit); \textbf{pass@$k$}, the fraction of tasks passing every check in at
least one of $k$ runtime executions of a fixed compiled program; and mean output tokens
per task. Direct generation compiles once and executes $k{=}3$ times, since a single
generation is never re-run. Agentic configurations already execute the program
repeatedly within their trajectory, so we score their final draft once, at $k{=}1$
(Appendix~\ref{app:harness}).

\paragraph{Models.} We evaluate five models: three frontier hosted models
(GPT-5.5, Claude Opus 4.8, Claude Sonnet 5), one smaller hosted model (GPT-5 mini),
and one open-weight model (Gemma-4 31B Instruction-Tuned). The direct-generation
leaderboard includes all five. Agentic configurations are run on Opus 4.8 and
Sonnet 5, the hosted models supported by the coding-agent CLIs we study.

\paragraph{Configurations.} Holding task, environment, and checks fixed, we vary the prompting context and whether the model may iterate against a compiler. \emph{Direct generation} is one shot with no tools: \textbf{zero-shot} supplies only the specification and any starting program, and \textbf{two-shot} adds two held-out examples; both use pass@3 to capture runtime variability. The remaining configurations are agentic and use pass@1, since they can already execute and revise code repeatedly. The \textbf{baseline} places the zero-shot prompt in a sandbox with a general-purpose coding agent that may read, edit, compile, and run the agent against the task's checks over multiple turns, but has no knowledge of AgentScript. Three \textbf{domain-skill packs} add generation, observation, and testing skills, differing only in how that guidance is authored and delivered: \textbf{v1} as inert documentation, \textbf{v2} as a live executable plugin, and \textbf{v3} as that same plugin with only the authoring skill's guidance rewritten (Appendix~\ref{app:skills}) . Finally, \textbf{afcode} is an alternative platform CLI with a built-in agent and a minimal tool set.

%% file: sections/03_evaluation.tex
\section{Results}
\label{sec:eval}

\begin{table}[!tb]
  \centering
  \caption{Pass rate is mean per-task partial credit;
  tokens are mean output per task}
  \label{tab:leaderboard}
  \label{tab:agentic}
  \small
  \begin{tabular}{llcccc}
    \toprule
    Configuration & Model & pass@$k$ (\%) & Compile (\%) & Pass rate (\%) & Tokens/task \\
    \midrule
    \multicolumn{6}{l}{\emph{Direct generation, zero-shot} ($k{=}3$)}\\
    & Claude Opus 4.8   & 25.0 & 38.0 & 31.3 & 9.9k \\
    & GPT-5.5           & 20.0 & 38.0 & 29.1 & 9.5k \\
    & Gemma-4 31B IT    & 15.0 & 48.0 & 28.1 & \textbf{7.0k} \\
    & GPT-5 mini        & 14.0 & 32.0 & 20.9 & 9.2k \\
    & Claude Sonnet 5   & 14.0 & 21.0 & 17.4 & 8.2k \\
    \midrule
    \multicolumn{6}{l}{\emph{Agentic}, Opus 4.8 ($k{=}1$); Sonnet 5 in Appendix~\ref{app:results}}\\
    Baseline (no skills)     & Opus 4.8 & 33.0 & 64.0 & 46.5 & 835k \\
    Skills v1 (inert docs)   & Opus 4.8 & 28.0 & 89.0 & 51.5 & 480k \\
    Skills v2 (live plugin)  & Opus 4.8 & 14.0 & 86.0 & 43.6 & 389k \\
    Skills v3 (rewritten)    & Opus 4.8 & \textbf{37.0} & \textbf{98.0} & \textbf{61.1} & 440k \\
    afcode (minimal CLI)     & Opus 4.8 & 21.0 & 89.0 & 45.3 & 293k \\
    \bottomrule
  \end{tabular}
\end{table}

\paragraph{Finding 1: models struggle, and the difficulty is concentrated in
authoring.} The best model reaches only $25.0\%$ pass@3 in a narrow
$14.0$--$25.0\%$ field. Even the strongest compile rate is under half ($48.0\%$), so most
failures never parse into valid AgentScript. Among programs that do compile many still
misbehave, Gemma-4 compiling $48.0\%$ of samples but solving only $15.0\%$.
Few-shots widen this gap rather than close it, lifting Opus's compile rate to
$60.0\%$ while \emph{lowering} its pass@3 to $21.0\%$ (Table~\ref{tab:twoshot}). By category (Table~\ref{tab:category}), \emph{every} model
scores $0.0\%$ on \textbf{authoring} and no configuration exceeds $16.7\%$, whereas
\textbf{optimization} is already $53.3\%$ zero-shot and \textbf{debugging} rises from
$22.5\%$ to $45.0\%$ under the strongest skill pack. That authoring collapses while
scaffolded editing does not points to generative coverage of the DSL, not local
reasoning, as the bottleneck; and no configuration wins all three categories.

\paragraph{Finding 2: scale and price are weak predictors.} Were capability tracking
scale and price, the largest hosted models would dominate every column; they do not.
Gemma-4 31B, the smallest and only open-weight model, has the highest compile rate of
any model ($48.0\%$ vs.\ $38.0\%$), the lowest token cost and a pass rate ($28.1\%$)
within three points of the best ($31.3\%$). A gap remains on
strict pass@3 ($15.0\%$ vs.\ $25.0\%$), so smaller models are not equal, but the
ordering is not the price ordering: among hosted models, Sonnet 5 trails the smaller
GPT-5 mini on both compile rate and pass rate. Defaulting to the most expensive model is
not clearly justified in this regime.

\paragraph{Finding 3: an agentic loop is the largest lever, and how domain guidance is
\emph{written} matters more than how it is \emph{packaged}.} Giving Opus a compiler to
iterate against raises pass@$k$ from $25.0\%$ to $33.0\%$ and increases mean pass
rate ($31.3\%\rightarrow46.5\%$): the largest gain from any harness change, driven by
the ability to compile, run, and revise its own program rather than framework
choice.

Adding skills on top makes the outcome sensitive to how that pack is authored. Moving from
inert docs (v1) to a live plugin (v2) changes delivery breadth and catalog together and
regresses below the skill-free baseline (Opus $28.0\rightarrow14.0\%$, Sonnet
$27.0\rightarrow16.0\%$), inviting the conclusion that executable delivery is the
culprit. It is not. Version~v3 holds delivery fixed and rewrites only the authoring
skill's guidance, halving its entry point ($830\rightarrow461$ lines) and splitting its
references for progressive disclosure at unchanged word count. That single edit takes
Opus to $\mathbf{37.0\%}$ pass@1, $98.0\%$ compile, and $61.1\%$ pass rate, the best
result in our study: packaging was never the binding constraint, guidance density was.
Under v2 the model spent more turns on failed than passed tasks ($15.4$
vs.\ $13.1$); under v3 mean turns fall to $12.0$.

%% file: sections/04_discussion.tex
\section{Discussion and Future Work}
\label{sec:discussion}

Current LLMs remain unreliable for DSLs authored by proprietary agents and larger models provide limited gains. The largest single lever is allowing the model to compile and run its own program; beyond that, the gains are uneven and must be measured rather than assumed. Our sharpest result concerns domain-skill packs, whose value depends less on packaging than on how their guidance is written: rewriting one skill for progressive disclosure improved strict accuracy by $+23$ points, exceeding gains from changes in delivery or catalog, suggesting that teams shipping skills to agents should treat guidance density as a design variable. Agentic loops and domain scaffolding largely overcome the DSL's surface-form difficulty, with compile rates reaching $98\%$, but behavioral correctness does not follow: even with compiler errors and graded results in hand, our best configuration leaves $40$ of $100$ tasks at partial credit. The bottleneck is therefore not the absence of correctness signals, but the ability to act on them, pointing to targeted repair from failing cases rather than broad regeneration~\citep{chen2023selfdebug,madaan2023selfrefine,shinn2023reflexion,chen2022codet,ni2023lever} as a promising next step.

%% file: sections/05_related_work.tex
\section{Related Work}
\label{sec:related}

\textbf{Code-generation benchmarks} target general-purpose languages: function
synthesis~\citep{chen2021evaluating,austin2021program}, competition
programming~\citep{li2022alphacode}, and repository-level issue
resolution~\citep{jimenez2023swe,yang2024sweagent}. \textbf{DSL and
low-resource generation} covers public, well-represented DSLs such as
text-to-SQL~\citep{yu2018spider} and IT-automation YAML~\citep{pujar2023ansible}, and
resource level across general-purpose languages~\citep{cassano2022multipl}; we target
instead a proprietary, in-production agent-authoring DSL absent from public data. \textbf{In-context
learning}~\citep{brown2020fewshot,dong2022icl,scholak2021picard,geng2023grammar} and
\textbf{execution-guided
self-correction}~\citep{chen2023selfdebug,madaan2023selfrefine,shinn2023reflexion,chen2022codet,ni2023lever}
are candidate mechanisms; our results point to the latter.
\textbf{Agents}~\citep{yao2023react} are here the artifact to be authored, not the
harness.

%% file: sections/06_appendix.tex
\section{The AgentScript language}
\label{app:language}

AgentScript is an indentation-structured declarative DSL. A program declares a
\texttt{system} block (persona and canned messages), a \texttt{config} block
(deployment identifiers), typed mutable \texttt{variables} that persist across a
conversation, and one or more \emph{topics}, each of which bundles the
\texttt{actions} the agent may invoke with natural-language \texttt{reasoning}
instructions that govern when to invoke them. Actions declare typed inputs and
outputs and bind those outputs back into conversation variables. Listing~1 shows
an abridged program in the shape the benchmark uses; identifiers are anonymized.

\begin{figure}[h]
\small
\begin{verbatim}
system:
    instructions: "You are an order tracking assistant for a retailer."
    messages:
        welcome: "Hi! Share an order number or email to get started."
        error:   "I can't reach the order system right now."

variables:
    customer_email: mutable string = ""
        description: "Customer's email address"
    customer_verified: mutable boolean = False
        description: "Whether customer identity has been verified"

start_agent order_locator:
    description: "Locates and validates customer order information"

    actions:
        Get_Customer_Info:
            description: "Retrieves customer information using email address"
            require_user_confirmation: False
            inputs:
                email: string
                    is_required: True
                    is_user_input: True
            outputs:
                customer_found: boolean
                customer_name:  string
            target: "flow://GetCustomerInfo"

    before_reasoning:
        if @variables.customer_email == "":
            set @variables.customer_verified = False
\end{verbatim}
\vspace{2pt}
{\small\textbf{Listing 1:} An abridged AgentScript program.}
\end{figure}

Two properties make the language a hard generation target. Semantics are carried by
fields whose names read as ordinary English but whose values change runtime behavior
(\texttt{require\_user\_confirmation} on a read-only lookup inserts a confirmation
turn), and the binding between action outputs and conversation variables is
positional and easy to permute without breaking compilation. Both yield programs
that compile and misbehave, which is the failure mode our results are dominated by.

\section{Benchmark construction}
\label{app:construction}

\paragraph{Provenance.} Tasks are seeded from real agent-development work. Each
\textbf{authoring} task pairs a natural-language specification with a
human-authored \emph{gold agent} that satisfies it. \textbf{Debugging} and
\textbf{optimization} tasks are built by injecting defects of known type into a
gold agent, so every task carries a recorded ground-truth fault, its location, and
its intended repair.

\paragraph{Composition.} $100$ tasks: $30$ authoring, $40$ debugging, $30$
optimization; $7$ are labelled basic, $22$ intermediate, and $71$ advanced. The
$40$ debugging tasks carry a median of $3$ injected defects each (max $5$) drawn
from eight recurring classes observed in real development, listed in
Table~\ref{tab:defects}. Optimization tasks inject a single behavioral defect into
a \emph{working} agent---the program compiles and runs, but violates a stated
objective (for example, a read-only lookup that pauses for user confirmation).

\begin{table}[h]
  \centering
  \caption{Defect classes injected into debugging tasks, with the number of tasks
  containing at least one defect of that class. Counts exceed $40$ because tasks
  carry multiple defects.}
  \label{tab:defects}
  \begin{tabular}{lc}
    \toprule
    Defect class & Tasks \\
    \midrule
    Reasoning lifecycle (hook ordering, side effects) & 35 \\
    Variable declarations (scope, shadowing, missing decls) & 30 \\
    DSL syntax (malformed blocks, indentation) & 25 \\
    Type serialization (input/output type mismatch) & 23 \\
    Config validation & 11 \\
    Locale / internationalization & 6 \\
    Voice channel & 5 \\
    Markdown rendering & 5 \\
    \bottomrule
  \end{tabular}
\end{table}

\paragraph{Acceptance checks.} Tasks ship $630$ executable test cases in total, a
median of $5$ per task (range $1$--$13$). A test case is a scripted multi-turn
dialogue (median $3$ turns, max $15$) together with the expected agent behavior at
its end: the actions the agent must have invoked, the conversation-variable state it
must have reached, and constraints on its response. Backend actions are stubbed with
fixed return values, so a dialogue is reproducible across runs and models. Every task
is scored all-or-nothing at the task level: a program passes only if \emph{all} of
its test cases pass, and a program that fails to compile scores zero.

\section{Prompts}
\label{app:prompts}

\paragraph{Zero-shot.} The default generation prompt in our codebase bundles a
syntax example into every call. Because that example is itself a demonstration, our
zero-shot condition removes it, so the model receives the specification and
constraints and no syntax help at all:

\begin{quote}\small\ttfamily
You are an expert AgentScript developer.\\[2pt]
Your task is to generate AgentScript code that meets this specification:\\[2pt]
\{specification\}\\[2pt]
Constraints:\\
\{constraints\}\\[2pt]
Return ONLY valid AgentScript code that satisfies the specification and compiles
cleanly. Do not include prose commentary, explanations, or a diff.
\end{quote}

\paragraph{Debugging.} Debugging tasks supply the broken program and ask for the
full corrected file, explicitly forbidding diffs and unrelated refactors so that
grading sees a complete program:

\begin{quote}\small\ttfamily
The following AgentScript file has a bug. Analyze the problem, identify the root
cause, and return the FULL corrected file.\\[2pt]
\{program\}\\[2pt]
Constraints the fix must satisfy:\\
\{constraints\}
\end{quote}

\paragraph{Optimization.} Optimization prompts supply the working program and the
observed symptom only. We deliberately \emph{omit} the constraints block here: for
these tasks the constraints spell out the intended repair, and including them would
leak the answer.

\paragraph{Two-shot.} The two-shot condition prepends two (specification, program)
demonstrations drawn from a held-out pool of gold agents that no benchmark task uses.

\paragraph{Agentic.} Agentic configurations receive the same zero-shot prompt seeded
into a sandbox containing the starting program (where the task supplies one) and helpers
for compiling and running the agent against the task's checks, so they may iterate on
both compiler errors and graded execution results. Direct generation receives neither.

\section{Harness and protocol}
\label{app:harness}

Each generated program is compiled in a real project environment on the host
platform; programs that compile are deployed and driven through each test case's
dialogue by a scripted user simulator, with stubbed backend actions. Agentic runs
execute in a per-task sandbox capped at $50$ assistant turns and $900$\,s of wall
clock, whichever comes first; the final \texttt{draft.agent} in the sandbox is what
gets graded, whether or not the agent announced completion.

\paragraph{Why $k$ differs by configuration.} Runtime execution is stochastic even for a fixed compiled program: the same program can produce different outcomes across runs. We therefore use different values of $k$ for direct LLM generation and agentic configurations to reflect the different opportunities each configuration has to respond to this runtime variability. For direct generation, the model produces a program in a single shot and does not observe or respond to its execution outcome. We therefore compile the generated program once and execute it three times, using pass@$k$ with $k{=}3$ to account for runtime variability. The task is counted as passed if any of the three executions clears every check. Thus, these three trials correspond to repeated executions of the same generated program, not to three independently generated programs.

Agentic configurations have a different evaluation protocol. The agent can compile and execute its program during its trajectory and use compiler errors, execution outcomes, and check results to revise the program over multiple turns. We therefore use pass@$k$ with $k{=}1$ and report the best result achieved during the agent's trajectory. Specifically, at each point where the agent produces or revises a program, we evaluate it against the task's checks, and the agent receives the corresponding feedback to guide subsequent revisions. The reported score is the outcome of the best program encountered during this process. This gives agentic configurations multiple opportunities to respond to runtime variability as part of their normal execution loop, whereas direct-generation configurations receive three independent executions of their single generated program.

\section{Additional results}
\label{app:results}

\begin{table}[h]
  \centering
  \caption{Two-shot direct generation ($100$ tasks; pass@3). Two-shot lift
  compile rate substantially for some models (Opus $38.0\rightarrow60.0$) but do not
  convert into correctness: only Sonnet 5 improves on pass@3, and by just two points (14.0→16.0); the strongest model drops four. Demonstrations teach syntax, not intent.}
  \label{tab:twoshot}
  \begin{tabular}{lcccc}
    \toprule
    Model & pass@3 (\%) & Compile (\%) & Pass rate (\%) & Tokens/task \\
    \midrule
    Claude Opus 4.8 & 21.0 & 60.0 & 30.9 & 14.4k \\
    GPT-5.5         & 20.0 & 36.0 & 25.9 & 11.0k \\
    Claude Sonnet 5 & 16.0 & 31.0 & 18.4 & 9.8k \\
    GPT-5 mini      & 15.0 & 24.0 & 19.1 & 10.7k \\
    Gemma-4 31B IT  & 14.0 & 47.0 & 28.4 & 7.0k \\
    \bottomrule
  \end{tabular}
\end{table}

\begin{table}[h]
  \centering
  \caption{Agentic configurations for both models (pass@1). The body table reports the
  Opus rows only; Sonnet is included here in full. v3 lifts Sonnet above v2 on every
  metric yet still trails v1's inert docs on strict pass@1.}
  \label{tab:agentic-full}
  \begin{tabular}{llcccc}
    \toprule
    Configuration & Model & pass@1 (\%) & Compile (\%) & Pass rate (\%) & Tokens/task \\
    \midrule
    Baseline (no skills)     & Opus 4.8 & 33.0 & 64.0 & 46.5 & 835k \\
    Baseline (no skills)     & Sonnet 5 & 20.0 & 65.0 & 40.4 & 1{,}563k \\
    Skills v1 (inert docs)   & Opus 4.8 & 28.0 & 89.0 & 51.5 & 480k \\
    Skills v1 (inert docs)   & Sonnet 5 & 27.0 & 79.0 & 43.7 & 1{,}125k \\
    Skills v2 (live plugin)  & Opus 4.8 & 14.0 & 86.0 & 43.6 & 389k \\
    Skills v2 (live plugin)  & Sonnet 5 & 16.0 & 62.0 & 29.4 & 1{,}123k \\
    Skills v3 (rewritten)    & Opus 4.8 & \textbf{37.0} & \textbf{98.0} & \textbf{61.1} & 440k \\
    Skills v3 (rewritten)    & Sonnet 5 & 23.0 & 91.0 & 49.5 & 914k \\
    afcode (minimal CLI)     & Opus 4.8 & 21.0 & 89.0 & 45.3 & 293k \\
    afcode (minimal CLI)     & Sonnet 5 & 28.0 & 79.0 & 52.7 & 400k \\
    \bottomrule
  \end{tabular}
\end{table}

\begin{table}[h]
  \centering
  \caption{Per-category pass@$k$ for the strongest model (Opus 4.8) across all
  configurations. Direct rows are pass@3; agentic rows are pass@1. Authoring resists
  every configuration (best $16.7\%$); optimization is already tractable zero-shot; no
  single configuration wins all three categories.}
  \label{tab:category}
  \begin{tabular}{lccc}
    \toprule
    Configuration & Authoring (\%) & Debugging (\%) & Optimization (\%) \\
    \midrule
    Zero-shot direct              & 0.0  & 22.5 & 53.3 \\
    Two-shot direct               & 10.0 & 15.0 & 40.0 \\
    Agentic baseline              & \textbf{16.7} & 25.0 & \textbf{60.0} \\
    Skills v1 (inert docs)        & 6.7  & 32.5 & 43.3 \\
    Skills v2 (live plugin)       & 13.3 & 15.0 & 13.3 \\
    Skills v3 (rewritten)         & 10.0 & \textbf{45.0} & 53.3 \\
    afcode (minimal CLI)          & 6.7  & 20.0 & 36.7 \\
    \bottomrule
  \end{tabular}
\end{table}

\section{Skill-pack versions}
\label{app:skills}

All three versions cover the same three capabilities---generation (author an agent
from a specification), observation (inspect an agent and localize faults), and
testing (execute an agent against checks). They differ as follows.

\textbf{v1 (inert docs)} is a curated three-skill subset copied into the sandbox as
static markdown. The agent may read the guidance; the accompanying hooks and scripts
do not execute.

\textbf{v2 (live plugin)} ships the same core skills inside the platform's real
agent plugin: $40$ skills, working hooks, helper scripts, a bundled language server,
and a platform CLI the agent may shell out to. Its authoring skill's entry point is
$830$ lines with $41$ reference files.

\textbf{v3 (rewritten guidance)} keeps v2's plugin shell, its $40$-skill catalog, its
tool policy, and its bundled language server unchanged. It rewrites only the authoring
skill: the entry point shrinks to $461$ lines ($-44\%$) and its references are split
from $41$ into $49$ finer, on-demand files, holding total reference word count
essentially flat ($88.3$k\,$\rightarrow$\,$89.0$k words). The change is a
progressive-disclosure restructure---a lean entry point that pulls in detail only when
a task needs it---not additional content.

Comparing v1 with v2 therefore varies delivery and catalog breadth together, and
cannot attribute the regression to either alone. Comparing v2 with v3 isolates
guidance: one file, restructured, everything else held fixed. Neither v2 nor v3 ever
invoked the bundled language server on any of the $100$ tasks---both validated through
the harness \texttt{compile.sh}---so v3's compile-rate gain cannot be attributed to a
new validation surface.